\documentclass[twocolumn]{aa}
\usepackage{graphicx}
\def\beq{\begin{equation}}
\def\eeq{\end{equation}}
\def\bey{\begin{eqnarray}}
\def\eey{\end{eqnarray}}

\def\pc{\, {\rm pc} }

\def\msun{M_\odot}

\def\lsun{L_\odot}

\def\vg{{\bf g}}
\def\vR{{\bf R}}
\def\grad{{\bf \nabla}}
\def\div{{\bf\nabla\cdot}}
\def\a0{$a_0$}

\begin{document}
\title{Roche Lobe Sizes in Deep-MOND Gravity}
\author{HongSheng Zhao\thanks{PPARC Advanced Fellow at University of St Andrews, 
Outstanding Young Overseas Scholar at Beijing Observatory}
}
\institute{.}
\offprints{H. Zhao, \email{hz4@st-and.ac.uk}} 
\date{}

\abstract{MOdified Newtonian Dynamics (MOND) is 
evolving from an empirical to a decent theory respecting  
fundamental physics after Bekenstein (2004) 
showed that lensing and Hubble expansion can be modeled rigourously 
in a Modified Relativity.  
The degeneracy of MOND with Dark Matter can be broken if we examine
the non-linear MONDian Poisson's equation in detail.
Here we study the effect of tides for a binary
stellar system or a baryonic satellite-host galaxy system.
We show that the Roche lobe is more squashed than the Newtonian case 
due to the anisotropic dilation effect in deep-MOND.
We prove analytically that the Roche lobe volume 
scales linearly with the ``true" baryonic mass ratio
in both Newtonian and deep-MOND regimes, 
insensitive to the modification to the inertia mass.
Hence accurate Roche radii of satellites can break 
the degeneracy of MOND and dark matter theory.
Globular clusters and dwarf galaxies of comparable luminosities
and distances show a factor of ten scatter in limiting radii; this
is difficult to explain in any ``mass-tracing-light" universe. 
The results here are generalizable to the intermediate MOND regime for a wide class of 
gravity modification function $\mu(g)$ (Zhao and Tian, astro-ph/0511754). 
}

%\pacs{PACS numbers : 95.35.+d; }

\maketitle
\section{Introduction}

The alternative gravity theory of Modified Newtonian Dynamics
(MOND) (Milgrom 1983) has been doing very well in fitting
kinematic data on galaxy scales, often better than the standard
cold dark matter theory.  Baryonic matter alone is sufficient to
account for the gravity in such theory. The predictive power of
this 20-year-old classical theory with virtually no free
parameters (Bekenstein \& Milgrom 1984) is recently highlighted by
the astonishingly good fits to contemporary kinematic data of a
wide variety of high and low surface brightness spiral and
elliptical galaxies; even the fine details of the ups and downs of
velocity curves are elegantly reproduced without fine tuning of
the baryonic model (Sanders \& McGaugh 2002, Milgrom \& Sanders
2003). Originally it was proposed empirically (Milgrom 1983) that
rotation curves of axisymmetric disk galaxies could be fit by an
acceleration $g \equiv \left| \vg \right|=V^2/r$ which is stronger
than the Newtonian gravity $GM/r^2$ by a spatially varying factor
$1/\mu$ in the weak regime defined by $g \le a_0 \sim 1.2
\times 10^{-8} {\rm cm}\sec^{-2}$; e.g., $\mu(g/a_0) =
\min(1,g/a_0) \le 1$. This empirical MOND relation can be elevated  
to a theory for an arbitrary baryon density distribution $\rho(\vR)$, where 
a curl-free gravity field $\vg=-\grad \Phi$ is the gradient of 
a conservative potential $\Phi(\vR)$ and satisfies an equation 
(Bekenstein \& Milgrom 1984)
\beq\label{binary} 
\div\left[ {\mu \vg \over 4\pi G} \right] = - \rho(\vR) \qquad
 \leftarrow \qquad \div {\epsilon {\bf E} \over 4\pi} = -e(\vR).
\eeq 
Here we made the analogy with the Poisson's equation of
a curl-free electric field ${\bf E}=-\grad \phi$ generated by a  
cloud of static eletrons of density $-e(\vR)$ 
in a medium with a spatially varying dielectric constant $\epsilon(E)$
if we map the field $\vg \leftarrow {\bf E}$, 
the density $\rho(\vR) \leftarrow e(\vR)$
and the factor ${\mu(g) \over G} \leftarrow \epsilon(E)$.  
The above modified Poisson's equation is actually a reformat of 
the Lagrange equation 
${\partial L \over \partial \Phi} = -\grad \cdot {\partial L
\over \partial \vg}$, where the Lagrangian is given by
\beq -L =  \!\! \int\!\! d{\bf R}^3 
\left[ \rho \Phi + {\bar{\mu} |{\bf g}|^2\over 8\pi G} \right], 
\qquad \vg(\vR) = - \grad \Phi(\vR), \eeq  
where we integrate over the source energy density plus 
the field energy density; 
the factor $\bar{\mu}(g) \equiv {\int \!\! \mu(g) d(g^2) \over g^2}$. This 
classical field theory formulation 
guarantees conservations of energy and (angular) momentum
of an isolated system.

In the past this non-relativistic formulation of MOND has been
criticized for being incomplete for modelling the bending of
light (but see Qin, Wu \& Zou 1995).  
This, too, has changed since its generalization into a
respectable relativistic theory (christened TeVeS by Bekenstein 2004), which 
includes Hubble expansion, and passes
standard tests to check General Relativity
(Skordis, Mota, Ferreira et al. 2005, Chiu, Ko \& Tian 2005); 
GR is merely one limitting case of TeVeS.

Nevertheless, a main challenge of working on MOND is its essential
subtle non-linearity and scale-dependency, which makes it
unreliable to extrapolate Newtonian intuitions. As a result, there are very
few predictions of MOND in the literature in dynamical situations
where the non-sphericity of the potential is essential. It is
encouraging that the recent work of Ciotti \& Binney (2004, see also Zhao 2005) shows
surprisingly simple analytical scaling relations exist even for
the highly non-linear and non-spherical two-body relaxation
problem in MOND. Here we show a surprisingly simple scaling of
tides or the Roche lobe of a binary system (on either stellar or
galaxy scales) if it is in the non-linear deep-MOND regime.  A
subtle difference from a naive Newtonian extrapolation is also
pointed out.  We compare the predicted Roche lobe sizes with the observed
limiting sizes of Milky Way satellites (globular clusters and dwarf galaxies)
of $10^{5-6.5}\lsun$.

\section{Roche lobe \& binary potential in deep-MOND}

One way to reach the deep-MOND regime so that $\mu(g)={g \over a_0}$
is to be at a distance $R$ 
sufficiently far way from an isolated galaxy of total baryonic
mass $M$ so that the gravity $g(R) \ll a_0$.
Here $g(R)$ and the spherical galaxy potential $\Phi_0(R)$
are approximately related to the Newtonian gravity $GM/R^2$ by 
\beq g(R) = {d\Phi_0(R) \over R}= \sqrt{GM a_0 \over R^2}, 
~~R \equiv \sqrt{z^2+y^2+x^2}. \eeq

Consider introducing a low-mass satellite of mass $m$ at a
position $(x,y,z)=(0,0,D_o)$ in the above galactic potential $\Phi_0(R)$.  
Following Milgrom (1986), we consider the effect of the self-gravity of a satellite
mass inside an external galaxy field $g(R) \sim \sqrt{GM a_0 \over R^2}$. 
The spatially slow-varying external field dominates the satellite gravity 
sufficiently far away from the satellite.  So the MOND ``dielectric index" 
$\mu \sim \mu(g) \sim g(R)/a_0 \sim \sqrt{G M \over a_0 R^2}$ 
varies very little in the vicinity of the satellite.  
To the first order in $m$ the perturbation in
potential is given by \beq\label{m'} \Phi_1 (x,y,z) = - {G  m' \over
\tilde{r}}, ~~m' \equiv {m \over \mu(g_{D_o})}, \eeq where $m'$ 
is the modified inertia of the satellite due to the external field $g_{D_o}=g(D_o)$,
and $\tilde{r}$ is the effective distance from the centre of the satellite
given by
 \beq \tilde{r}=\!\!\sqrt{(z-D_o)^2 + \left(y^2+x^2\right) (1+\Delta)},~~
\Delta \equiv \left.{d\ln \mu \over d\ln g}\right|_{g=g(D_o)} \!\!\!\!\!\!\!\!
\label{dilate} \eeq 
where $1+\Delta(g)$ is a shape factor; \footnote{The ``dielectric index" 
$\mu({|{\bf g}| \over a})$ is more sensitive to perturbation along
the external field $g(D_o)\hat{z}$ direction than perpendicular because 
$|{\bf g}+d{\bf g}|=\left[(g_{D_o}+d g_z)^2
+(d g_x)^2 + (d g_y)^2\right]^{1 \over 2}$ depends on the perturbation 
$d g_z$ to first order, and $d g_x$ and $d g_y$ to second order.}
clearly $1+\Delta(g)=2$ in the deep-MOND regime where $\mu \sim g/a_0$, and $1+\Delta(g)=1$ for strong gravity.

The above formulation allows us to approximate the potential of,
e.g., the Milky Way galaxy with a satellite.  Substitute in the
expressions for $\mu$, $g(R)$, $R$ and $\tilde{r}$, the combined
potential is then given by 
 \beq
\Phi=\Phi_0 (R)+ \Phi_1 =  (GMa_0)^{1 \over 2} \ln\!\!\sqrt{z^2+y^2+x^2}
    - {G m' \over \tilde{r}},
 \eeq
which is an axisymmetric prolate potential with
two centres separated by distance $D_o$ along the z-axis, where the
two terms represent the MONDian potential of the Milky Way and
perturbation due to the satellite; here
 \beq\label{mrprime} 
m' = m \sqrt{D_o^2 a_0 \over G M},  \qquad \tilde{r}=\sqrt{(z-D_o)^2+ 2 y^2+ 2 x^2}.
 \eeq 
 We also note that the density $\rho=0$
along the z-axis for the gravity field ${\bf g}=-\grad \Phi$ of
the above combined potential; this can be verified by evaluating
the modified Poisson's equation $\rho=- \grad
\cdot {{\bf g}\mu \over 4 \pi G}$.

\begin{figure}{}
\resizebox{9cm}{!}{\includegraphics{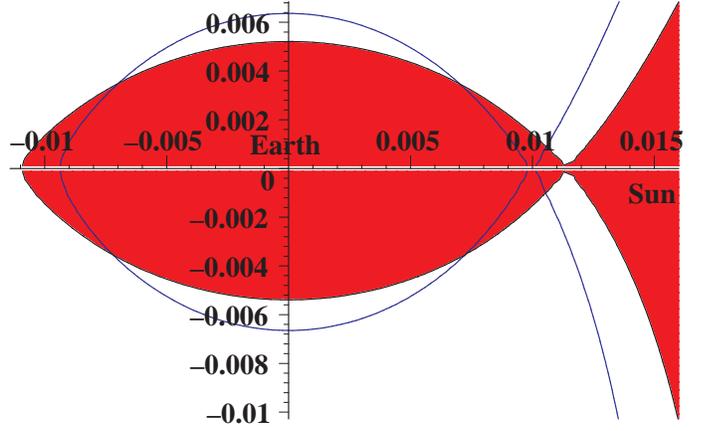}} \caption{ shows the
re-scaled Roche lobes (contours of the effective potential) in the
equatorial $xz$ plane (lower half) and in the vertical $yz$ plane
(upper half) of a hypothetical isolated Earth-Sun binary with a
mass ratio $3\times 10^{-6}$ in the strong gravity regime (say
with the separation $D_o=1$AU, thin blue lines), and in the weak gravity
regime (say with separation $D_o=0.1$pc and $G\msun D_o^{-2}a_0^{-1} \sim
0.1$, shaded areas). The Earth is at origin and the Sun
is at unit length to the right. The inner Lagrangian point is a
saddle point between the Earth and the Sun, which is slightly
further away from the Earth in the deep-MOND regime than in strong
gravity regime. }\label{lobe}
\end{figure}

Let the low-mass satellite with $m/M \ll 1$ rotate around the
galaxy centre (fixed) with an angular velocity $\Omega_0 \hat{\bf
y}$, then particles in the corotating frame conserve the Jacobi
energy with an effective (triaxial) potential \beq\label{omega}
\Phi_e(x,y,z) \equiv \Phi(x,y,z) - {x^2 + z^2 \over 2}\Omega_0^2,
~~ \Omega_0 = {\sqrt{G M a_0} \over D_o}. \eeq

The inner or outer Lagrangian points is then calculated from the
saddle point of the effective potential where 
 \beq \left.{\partial \Phi_e(0,0,z)
  \over \partial z} \right|_{z=D_o \pm r_L} \!\!\!\!\!\!\!\!\!\!\!\!\!\!\!\!\! =0 = 
  {\sqrt{G M a_0} \over D_o \pm r_L} \pm {G m' \over r_L^2} - \Omega_0^2 (D_o \pm r_L)
\eeq which defines the Lagrange radius $r_L$.
Taylor-expand the above to first order in $r_L/D_o$
(the 0th term cancels due to eq.~\ref{omega} and eq.~\ref{mrprime}), we
have  
 \beq\label{lag} \left({m \over r_L^3}\right)
 \left({M \over D_o^3}\right)^{-1} \!\! = 1+ \zeta \equiv \left[1 + { \Omega_0^2 D_o^2 \over \sqrt{G M a_0}}\right] =2.
 \eeq 
So inside the Lagrange radius the average
density of the satellite equals twice the average density inside
the orbit of the satellite in the weak gravity regime.  Note that
the masses here $m$ and $M$ are true baryonic masses of the
binary, {\it not} the modified inertia masses\footnote{Dimensional analysis
only cannot tell whether the dimensionless ${r_L \over D_o}$ scales like 
$\left( {m \over M}\right)^{n}$ or $\left({m' \over M}\right)^{n}$, 
where $m' ={m \over \mu}$ is the modified inertia (cf. eq.~\ref{m'}).}. 
The scaling that $r \propto m^{1/3}$ 
is also confirmed by the numerical simulations of Brada \& Milgrom (2000).

The shape of the Roche lobe is defined by the contour of the
effective potential (eq.~\ref{omega}) passing through the Lagrange
point.  Finding the roots analytically yields (Zhao \& Tian 2005)  
\beq r = \left[ 1,{\sqrt{2} \over 3},
(\sqrt{10}+\!\!\sqrt{2})^{1 \over 3} \!\! -(\sqrt{10}-\!\!\sqrt{2})^{1 \over 3} \right]
\left({m \over 2 M}\right)^{1 \over 3}\!\! D_o, \eeq
which are intersections with the long z-axis, intermediate x-axis and rotation
y-axis respectively.  
Of the three radii, the Intermediate Roche (IR) radius compares
most directly with observed size in the sky plane for a distant satellite, and is given by
\beq\label{axis}
{r_{\rm IR} \over D_o}\left({m \over M}\right)^{-{1 \over 3}} = 
{\sqrt{2} \over 3} \left({1 \over 2}\right)^{1 \over 3}= 0.374.
\eeq
Projection effects make the observed radius in between the short semi-axis and 
the long semi-axis of the Roche lobe.  Hence the intermediate axis ($r_{\rm IR}$ instead of $r_L$) 
is the best compromise among the three to approximates the observed size.

\section{Observed instantaneous Roche lobe}
 
If MOND is correct the Roche lobe would act as Nature's balance to 
weigh the relative baryonic content of a secondary vs. a primary star, 
or a satellite vs. its host galaxy.  Interestingly the Roche lobe 
satisfies the same scaling relation 
${r_{\rm IR} \over D_o}\left({m \over M}\right)^{-1/3} = cst$, 
but the $cst=0.462$ in strong gravity regime (Binney \& Tremaine 1987) 
while $cst=0.374$ in deep-MOND.  E.g., in a gedanke experiment where 
we take the solar system out of the Galaxy, and 
increase the Earth-Sun distance from 1AU to 0.1pc
(the separation of the widest known binary stars) so that near the
inner Lagrangian point of the system the gravity drops from the
strong regime to the weak regime. 
Fixing the Earth-Sun mass ratio $m/M=3\times 10^{-6}$, 
the rescaled intermediate Roche lobe radius 
$r_{\rm IR}/l$ should decreases slowly by a subtle amount from
$0.462 \times (m/M)^{1/3} = 0.0067$ radian (for strong gravity) to 
$0.374 \times (m/M)^{1/3} = 0.0054$ radian (for weak gravity); 
cf. eq.~(\ref{lag}) and~(\ref{axis}) and see Fig.~\ref{lobe}.
Likewise the aspect ratios of the Roche lobe evolves from 
$1:2/3:9^{1/3}-3^{1/3}=1:0.667:0.638$ (Binney \& Tremaine 1986)
to about $1:0.471:0.456$, and the volume of Roche lobe evolves from
$\sim {4 \pi D_o^3 \over 3} {m \over 7 M}$ to 
$\sim {4 \pi D_o^3 \over 3} {m \over 9 M}$.  
The Roche lobe is more squashed in MOND than in Newtonian gravity
(cf. Fig.1).

Interestingly, the same rescaled Roche radius can be predicted if we
substitute the Earth-Sun binary by a satellite (either a dwarf
spheroidal or a globular cluster) of a typical luminosity $\sim
3\times 10^5\lsun$ orbiting a luminous host galaxy of $\sim
10^{11}\lsun$ so that the baryonic mass ratio is about Earth-Sun
mass ratio.  The self-gravity around an extended object of mass
distribution $m(r)$ becomes weak compared to $a_0$ outside a
radius 
\beq r^{\rm w}\! =\!\!\sqrt{Gm(r) \over a_0}=\cases{0.1\pc & $m=10\msun$, \cr
10\pc & $m=10^5\msun$, \cr 
10^4\pc & $m=10^{11}\msun$.\cr}\eeq
Consider globular clusters and dwarf galaxies of the Milky Way
much further than $10{\rm kpc}$.  The outer envelope (well outside 10pc) of these
objects are generally in {\it the mildly-weak to the deep-MOND regime}. 
A satellite is generally on a non-circular orbit, nevertheless, 
an instantaneous Roche lobe radius can still be defined by the $r_{\rm IR}$ 
as if the satellite is orbiting on a circular orbit at its present orbital radius $D_o$
(approximately the distance from the Sun $D$) for an outer halo satellite.    
So we can rewrite eq.~\ref{axis} as
\beq
{r_{\rm IR} \over D} = A  \left({L_{\rm sat} \over L_{\rm MW}}\right)^{{1 \over 3}},
\qquad A \equiv 0.374 {D_o \over D}  \sim 0.374,
\eeq
where in estimating the instantaneous Roche radius $r_{\rm IR}$ (cf. eq.~\ref{axis}) 
we have assumed satellites have identical mass-to-light ratio as the Milky Way.   
This suggests comparable sizes $r_{\rm IR}$ for distant satellites 
at similar present distances $D$ or $D_o$ and comparable luminosity $L_{\rm sat}$.  

The actual direct observable is the limiting angular size $\theta_{\rm lim}$ of a satellite
seen from the Sun's perspective.  If a satellite fills the MONDian Instantaneous Roche Lobe,
we expect to observe an angular size $\theta_{\rm lim}=r_{\rm IR}/D$ perpendicular to the line of sight.
From these observables we can construct an observable ``filling factor" 
\beq\label{etaobs} F_{\rm lim} \equiv {\theta_{\rm lim} \over r_{\rm IR}/D} =
{\theta_{\rm lim} \over A} \left({L_{\rm sat} \over L_{\rm MW}}\right)^{-{1 \over 3}}.
\eeq
In the real world, the limiting radius should be somewhat smaller than the instantaneous Roche radius, i.e, 
$F_{\rm lim} \le 1$
because (i) the satellite likely remembers the truncation set by the stronger tide at some 
smaller orbital radius between last pericentre $r_p$ and present distance $D$, 
depending on whether a satellite can relax quickly in one orbital time (Bellazzini 2004);
(ii) this truncation might be beyond 
the limit of reliable observation $r_{\rm lim}$ simply because we run out of bright stars 
(Grebel, Odenkirchen \& Harbeck 2000).
Many satellites (cf. Dinescu, Keeny, Majewski et al. 2004 and references therein) 
are on nearly circular orbits (Ursa Minor and Canis Major dwarves) or are presently within
a factor of two to their pericentres (Pal3, Pal13, Fornax, Sgr
and LMC).  Theoretically a particle is typically found at the
geometric mean $\sqrt{r_p r_a}$ of its pericentre and apocentre
in a logarithmic potential.  So $D \sim D_o \sim (1-2) r_p$
typically even for a very radial orbit with $r_p:r_a \sim 1:4$, 
i.e., we expect a mild scatter of $F_{\rm lim}$ between $0.5-1$, 
allowing for radial orbits.  The filling factor should be nearly unity for globular clusters
since many galactic and some extragalactic globulars
(Harris, Harris, Holland et al. 2002) apparently fill their Roche lobes judging from 
extra-tidal stars in power-law profiles revealed
by deep observations wherever available (e.g., Leon, Meylan \& Combes 1996).

\begin{figure}{}
\resizebox{9cm}{!}{\includegraphics{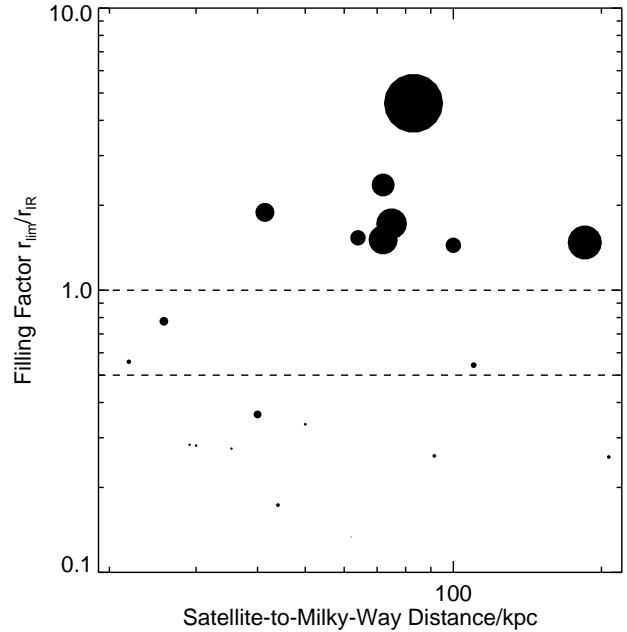}} \caption{ shows the MONDian ``filling factor" 
for outer Galactic satellites (20-200kpc), i.e., the ratio of
the observed limiting radius $r_{\rm lim}$ and the instantaneous Intermediate Roche radius $r_{\rm IR}$
(cf. eq.~\protect{\ref{etaobs}}, with a nominal luminosity $L_{\rm MW}=4\times 10^{10}\lsun$ for the Galaxy).  
The overall distribution is much broader than 
the range expected in MOND (indicated between the two thick dashed lines), and 
many are outliers with $r_{\rm lim}>r_{\rm IR}$. The
symbol sizes are proportional to the apparent limitting sizes ${r_{\rm lim} \over D}$ of
satellites (the smallest/biggest symbols correspond to physical radii of 40pc/4000pc, 
observational errors on distance and $\theta_{\rm lim}$ are 10\% typically), 
and the sample includes both globular clusters and
dwarf galaxies with luminosities between $10^5-10^{6.5}\lsun$
(data from Harris 1996, Mateo 1998).
}\label{scatter}
\end{figure}

These expectations, however, are not borne out by Fig.~\ref{scatter}, which shows 
surprisingly large scatter of the observed $F_{\rm lim} \sim 0.1-5$ for
distant Milky Way satellites with comparable luminosities ($10^5-10^{6.5}\lsun$). 
It is also difficult to understand why our expectation $F_{\rm lim} \le 1$ (cf. eq.~\ref{etaobs}) 
is contradicted most strikingly by systems of larger tidal radii (larger symbols).  
Surely observations carry errors.  
The distance factor $A$ is insensitive to the typical 10\% distance error.
Satellites often change profiles at $\theta_{\rm lim}$, so $\theta_{\rm lim}$ 
is well-defined with very little error.
Finally many satellites are in mild MOND regime with a Roche lobe size more rigorously given by (Zhao \& Tian 2005)
\beq
{{\rm Intermediate~Roche~Size} \over {\rm Orbital~Radius}} = {2 \over 3\sqrt{1+\Delta}} \left({1+\Delta \over 3+\Delta}{m \over M}\right)^{1 \over 3},
\eeq  
which is a smooth interpolation of eq.~(\ref{axis}) in deep-MOND and the familiar Newtonian prediction
(Binney \& Tremaine 1987) as the dilation factor (cf. eq.~\ref{dilate}) $1+\Delta({g \over a_0})$ changes 
from $2$ to $1$.  For any $\mu$-function
of the Milky Way (cf. Famaey \& Binney 2005), this correction is at most only 20\% 
(the factor 0.374 in eq.~\ref{axis} changed to 0.462).

In short, it is likely challenging for any theory of structure
formation of a baryonic MOND universe to explain the puzzling
large scatter in the rescaled Roche radius without fine-tuning of
satellite orbits and mass-to-light ratios.  It is less challenging for 
dark matter theories; $10^5$-star satellite objects could form either inside
or outside a small dark halo.  The scatter of satellite sizes echos with 
similar scatter of Einstein ring sizes around high-redshift lens galaxies 
(see lensing models of Zhao, Bacon, Taylor \& Horne 2005), highlighting
possible difficulties of mass-trace-light models.

%\vfill\eject

\begin{acknowledgements}
HSZ thanks the referee for constructive comments, 
LanLan Tian and Huanyuan Shan for help.  This work is
partial supported by Chinese NSF grant 10428308.
\end{acknowledgements}

%\vfill\eject

\end{document}